# Creation of a neutrino laboratory for search for sterile neutrino at SM-3 reactor

A.P. Serebrov[a*], V.G. Ivochkin[a], R.M. Samoylov[a], A.K. Fomin[a], V.G. Zinoviev[a], P.V. Neustroev[a],
V.L. Golovtsov[a], N.V. Gruzinsky[a], V.A. Solovey[a], A.V. Cherniy[a], O.M. Zherebtsov[a],
V.P. Martemyanov[b], V.G. Zinoev[b], V.G. Tarasenkov[b], V.I. Aleshin[b], A.L. Petelin[c], S.V. Pavlov[c],
A.L. Izhutov[c], S.A. Sazontov[c], D.K. Ryazanov[c], M.O. Gromov[c], V.V. Afanasiev[c], L.N. Matrosov[a],
M.Yu. Matrosova[a]

[a]*Petersburg Nuclear Physics Institute, Gatchina, 188300 Russia*

[b]*NRC «Kurchatov institute», Moscow, 123182 Russia*

[c]*JSC "SSC RIAR", Dimitrovgrad, 433510 Russia*

**Abstract**

In connection with the question of possible existence of sterile neutrino the laboratory on the basis of SM-3 reactor was created to search for oscillations of reactor antineutrino. A prototype of a neutrino detector with scintillator volume of 400 l can be moved at the distance of 6-11m from the reactor core. The measurements of background conditions have been made. It is shown that the main experimental problem is associated with cosmic radiation background. Test measurements of dependence of a reactor antineutrino flux on the distance from a reactor core have been made. The prospects of search for oscillations of reactor antineutrino at short distances are discussed.

**Introduction**

At present there is a widely spread discussion of possible existence of a sterile neutrino having much less cross-section of interaction with matter than, for example, reactor electron antineutrino. It is assumed that owing to reactor antineutrino transition to sterile condition, oscillation effect at a short reactor distance and deficiency of a reactor antineutrino beam at a long range are likely to be observed [1, 2]. Moreover, sterile neutrino can be regarded as being a candidate for the dark matter.

Ratio of the neutron beam observed in experiments to the predicted one is estimated as 0.927±0.023 [1]. The effect concerned comprises 3 standard deviations. This is not yet sufficient to have confidence in existence of the reactor antineutrino anomaly. The method of comparing the measured antineutrino beam with the expected one from the reactor is not satisfactory because of the problem of accurate estimation of a reactor antineutrino beam and efficiency of an antineutrino detector.

The idea of oscillation can be testified by direct measuring of a beam variation effect and antineutrino spectrum at a short reactor distance. A detector is supposed to be movable as well as spectrum sensitive. Our experiment focuses on the task of either confirming at a certain confidence level a possible existence of sterile neutrino or disproving it. For seeking oscillations to sterile neutrino, it is necessary to register variations of the reactor antineutrino beam. If such a process does occur, it can be described by an oscillation equation:

$$P(\tilde{\nu}_e \to \tilde{\nu}_e) = 1 - \sin^2 2\theta_{14} \ \sin^2(1.27\frac{\Delta m_{14}^2[eV^2]L[m]}{E_{\tilde{\nu}}[MeV]}) \qquad (1),$$

where $E_{\tilde{\nu}}$ is antineutrino energy, with oscillations parameters $\Delta m_{14}^2$ and $\sin^2 2\theta_{14}$ being unknown. For the experiment to be conducted, one needs to make measurements of an antineutrino beam and spectrum at short distances 6-12 m from, practically, a point source of antineutrino.

We have studied possibility of making new experiments at research reactors in Russia. It is research reactors that are required for performing such experiments as they possess a compact reactor core center and can be situated at a sufficiently small distance from possible location of a neutrino detector. Unfortunately, the research reactor beam hall has a fairly large background of neutron and gamma quanta, which makes it difficult to perform low background experiments. Due to some peculiar characteristics of its construction, reactor SM-3 provides the most favorable conditions for conducting an experiment on search for neutron oscillations at short distances.

### 1. Reactor SM-3

Initially 100-MW reactor SM-3 was designed for carrying out both beam and loop experiments. Five beam halls were built, separated from each other with big concrete walls as wide as ~1 m (Fig. 1). This enabled to carry out experiments on neutron beams, without changing background conditions at neighboring installations. Later on, the main experimental program was focused on the tasks concerned with irradiation in the reactor core center. For 25 years a sufficiently high fluence had been accumulated on materials of the reactor cover, which necessitated its replacement. Setting a new reactor cover into an old reactor tank was the simplest decision to be taken. Such a decision, however, resulted in raising a reactor core center 67 cm higher than the previous position.

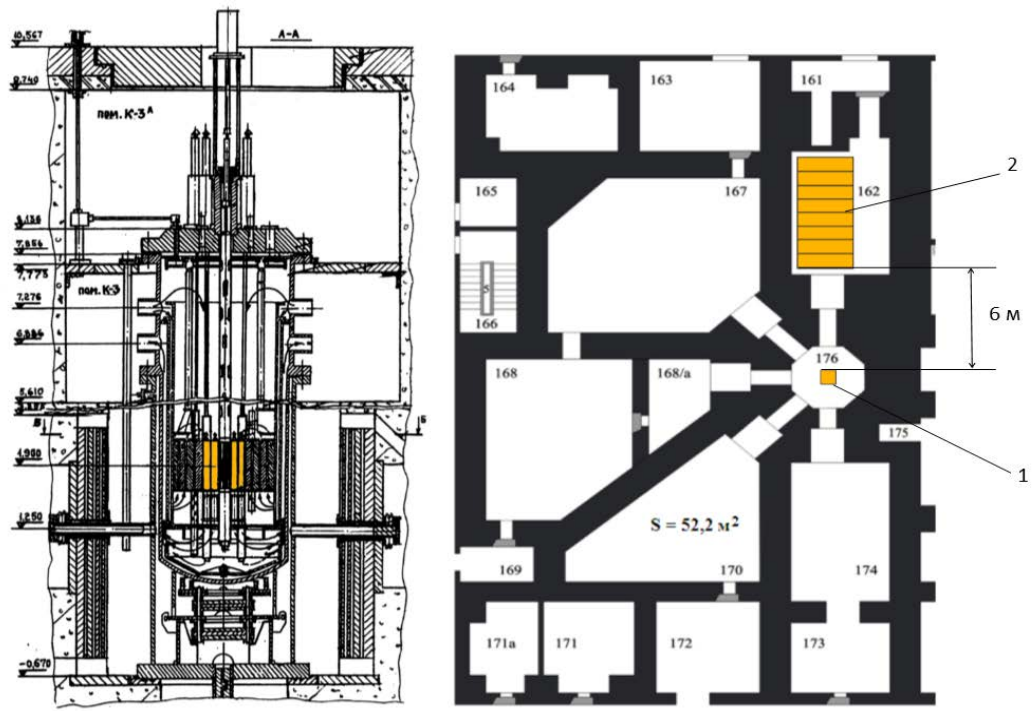

Fig. 1. Detector location at reactor SM-3: 1 – reactor core, 2 – antineutrino detector.

Horizontal beam channels were sacrificed for the sake of priority of loop experiments. Neutron flux in the location place of the former beam channels was decreased by four orders of value. Respectively, it caused decrease of neutron background in the former beam halls, which became about $4 \cdot 10^{-3}$ n/(cm$^2$s) (on thermal neutrons). It is approximately by 4–5 orders of value lower than a typical neutron background in the beam hall of a research reactor. Lately in making preparations for an experiment on search for transitions of reactor antineutrino to sterile state by reactor SM-3, upgrading of slide valve of the former neutron beam has been completed. As a result, the background of fast neutrons has dropped to the level of a few units by $10^{-3}$ n/(cm$^2$s), i.e. practically, to the level of neutron background on the Earth surface caused by cosmic radiation. These conditions are most preferable for a neutron experiment to be performed. Other advantages of SM-3 reactor are a compact reactor core center (35×42×42 cm) with high reactor power being equal to 100 MW, as well as a sufficiently short distance (5 m) from the center of a reactor core to the walls of an experimental hall. Besides, of special significance is the fact that an antineutrino beam can be measured within a sufficiently wide range from 6 to 13 meters. Up to $1.8 \cdot 10^3$ neutrino events are expected to occur per day at the reactor power 100 MW, at the distance of 6 m from a reactor core, in the volume of 1 m$^3$.

## 2. A neutrino detector model and its testing by WWR-M reactor.

With the view of preparation for the experiment «Neutrino-4» by SM-3 reactor [3-6] we did an experimental research with a detector model «Neutrino-4» at WWR-M reactor. An experimental task was to register reactor antineutrino in the conditions of high background of cosmic radiation on the surface of the Earth as well as in the conditions of neutron and gamma background in an experimental hall of a research reactor. This experiment was to investigate principal possibility of performing such an experiment by SM-3 reactor. The experiment demonstrated that of main difficulty is the background of cosmic radiation producing correlated events which are difficult to distinguish from those of reactor antineutrino recordings. Neutron and gamma background in an experimental hall can be suppressed by 4-5 orders of value with passive shielding made of lead, borated polyethylene and concrete. It is to be noted that the most optimal sequence of placing protection layers is as follows: concrete is to be located outside, followed by lead and then borated polyethylene inside. An internal layer of borated polyethylene is absolutely necessary, as it provides protection from neutrons emitted on lead by muons. Moreover, this process is likely to give rise to correlated events. After testing the detector model was transported to SM-3 reactor.

The detector model scheme «Neutrino-4» is shown in Fig. 2. The detector volume 0.9x0.9x0.5 m$^3$ is filled with liquid scintillator with addition of Gd. The detector makes use of 16 photoelectron multipliers PMT-49b located on the upper surface of the detector. The scintillation type detector is based on using the reaction $\tilde{\nu}_e + p \rightarrow e^+ + n$. At the first moment the detector registers positron, whose energy is determined by antineutrino energy and also registers 2 annihilation gamma quanta with energy 511 keV each. At the second moment neutrons emerging in reaction are absorbed by Gd to form a cascade of gamma quanta with total energy about 8 MeV. The detector keeps records of two subsequent signals from positron and neutron - so named correlated events.

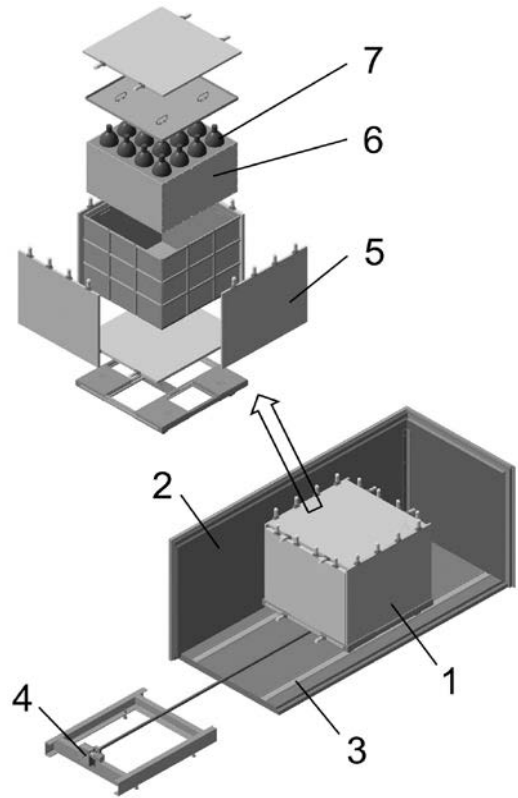

Fig. 2. Detector layout: 1 –detector of reactor antineutrino, 2 – passive shielding made of lead (6 cm) and borated polyethylene (16 cm), 3 – rails, 4 – device for detector displacement, 5 – plates of active shielding , 6 –liquid scintillator , 7 –PMT.

The antineutrino spectrum is restored from that of positron, since in the first approximation the relationship between positron energy and that of antineutrino is linear: $E_{\tilde{\nu}} = E_{e^+} + 1.8$ MeV. Scintillator material is made up of mineral oil with added Gd 1 g/l. Light output of scintillator BC-525 is $10^4$ photons for 1 MeV. The detector is surrounded by 6 scintillation plates as big as 0.9x0.9x0.03 m, with photoelectron devices being anticoincidence shielding from cosmic muons. After conducting test experiments at WWR-M reactor, research on a neutron detector model was carried out at SM-3 reactor, where by that time a neutron laboratory and passive shielding of the detector had been prepared.

### 3. Passive shielding of a neutron detector at SM-3 reactor

Layout of passive shielding from the outside and inside is given in Fig. 3. It is created from elements based on steel plates 1 x 2m, 10mm thick, to which are attached 6 sheets of lead as thick as 10 mm. The cabin volume is 2x2x8 m. From the inside the cabin is covered with plates of borated polyethylene 16 cm thick. The total weight of passive shielding are 60 tons, the volume of borated polyethylene is 10 m$^3$. Inside passive shielding there is a platform with the antineutrino detector which can be moved with a step motor along the rails within the range of 6 to 12 meters from the reactor core center. A neutrino channel can be entered by means of a ladder through the roof with the removed upper unit, as shown in Fig. 3. Loading of the detector into a neutron channel is carried out from the main hall through a trap door in the building ceiling. In this case an overhead crane of the main hall is used.

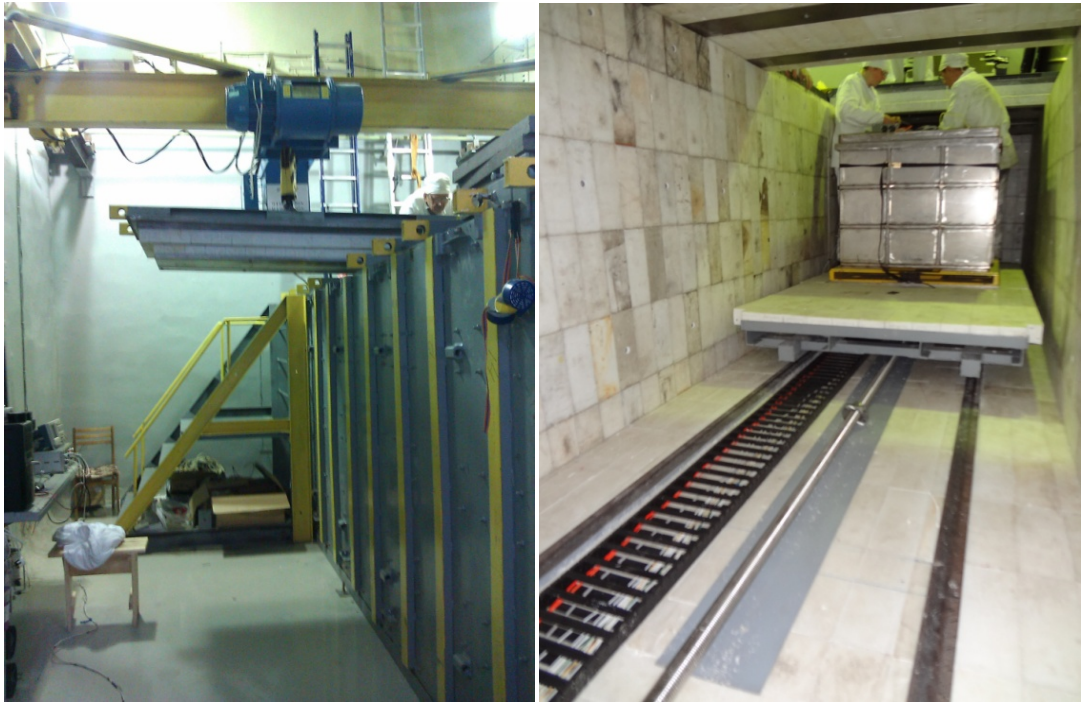

Fig. 3. General view of passive shielding: from the outside and inside. The range of detector dislocation is 6 - 12 m from the reactor core center.

## 4. Investigation of background conditions with gamma and neutron detectors for an experiment «Neutrino-4» at SM-3 reactor

### 4.1 Background of gamma rays in a neutrino laboratory and inside passive shielding of a neutrino detector

For measuring gamma ray spectrum a detector NaJ(Tl) 60×400 mm was applied. It has higher sensitivity due to the crystal length of 400 mm. Spectrum of signals of such a detector is given in Fig. 4. It demonstrates $^{60}$Co and $^{137}$Cs radioactive contamination and muon peak of cosmic radiation. In switching on the reactor, spectrum shows count of gamma-quanta from neutron capture in an iron-concrete shielding of the reactor. During a reactor operation the background spectrum within energy range from 3 MeV to 8 MeV contains in its composition an instant gamma-radiation emerging at interaction of thermal neutrons with iron nuclei in constructive materials. Within this energy range the detector count intensity at the reactor on is higher than that at the reactor off by factor of 22. It is this energy range which is of great importance, since it corresponds to gamma-quanta energy at neutron capture by Gd.

Gamma-radiation of isotopes of $^{137}$Cs, $^{60}$Co is independent of the reactor operation mode and is caused by radioactive contamination from the building floor and walls. In spite of such measures as pouring the concrete floor with an iron grit and reconstruction of slide valve which reduced gamma radiation background in this energy range 5-6 times, it does remain high enough, which confirms necessity of passive shielding for a detector from gamma quanta.

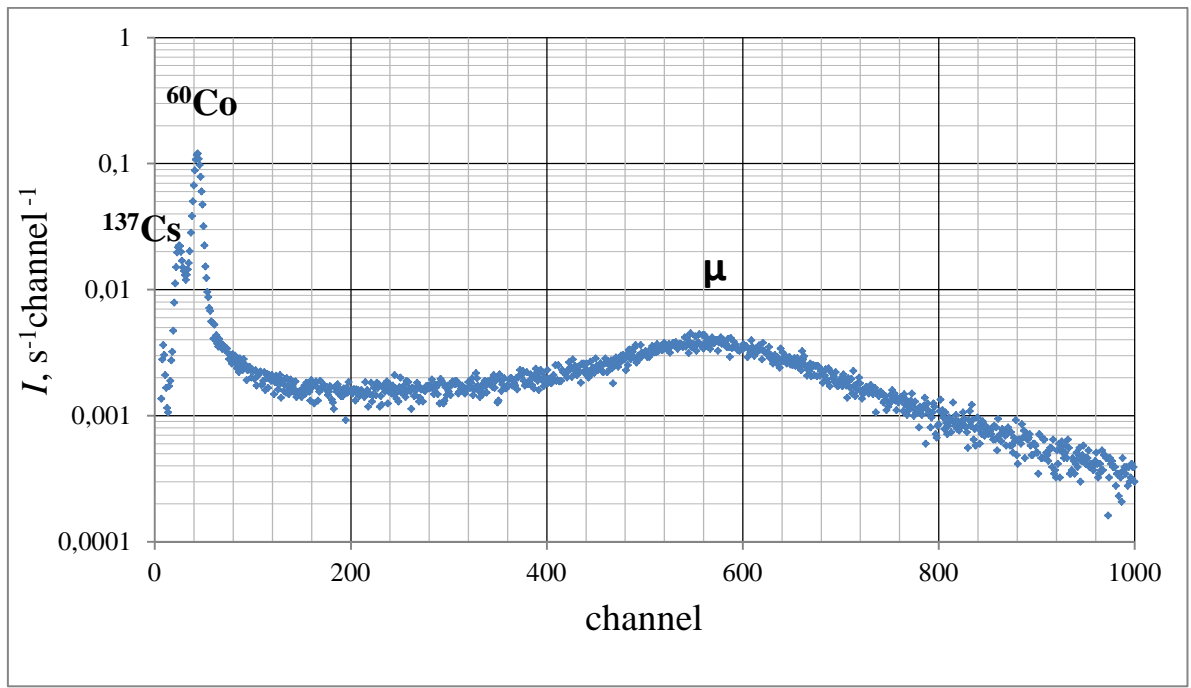

Fig. 4. Background spectrum with NaJ detector in a neutron laboratory beyond passive shielding.

In order to estimate possibility of suppressing gamma radiation background at an operating reactor, one recorded gamma radiation spectra in shielding of lead bricks (the wall 5 cm thick) and without it. The detector NaJ (60x400 mm) was placed on the floor of the platform for dislocating a neutron detector at the distance of 4 m from the gate device of a neutron channel. Spectra and results of gamma radiation suppression with the lead shielding at an operating reactor are shown in Fig. 5.

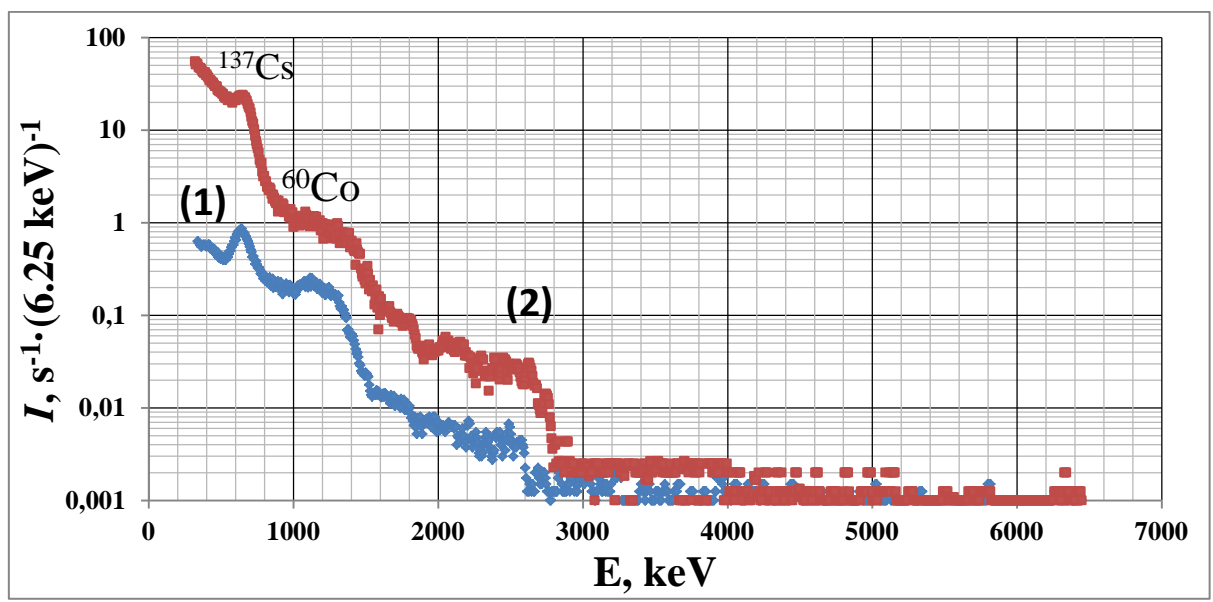

Fig. 5. Spectra of background gamma-radiation at the lead shielding (1) and without it (2), recorded with the detector NaJ (60x400 mm). Reactor power is 90 MW.

Within energy range from 1440÷7200 keV (from $^{40}$K and higher), 5 cm of lead shielding makes the level of background gamma radiation 4.5 times lower, which proves that its creation for the detector is justifiable. However, it is to be noted that neutron background resulting from the interaction of cosmic muons with lead nuclei increases inside the lead shielding. Indeed, the same (5 cm) lead shielding around the neutron detector resulted in raising its count rate by two times. Thus, inside the lead shielding there must be located another one made from borated polyethylene. Investigations, carried out with NaJ detector before and after installing passive shielding are given in Fig. 6.

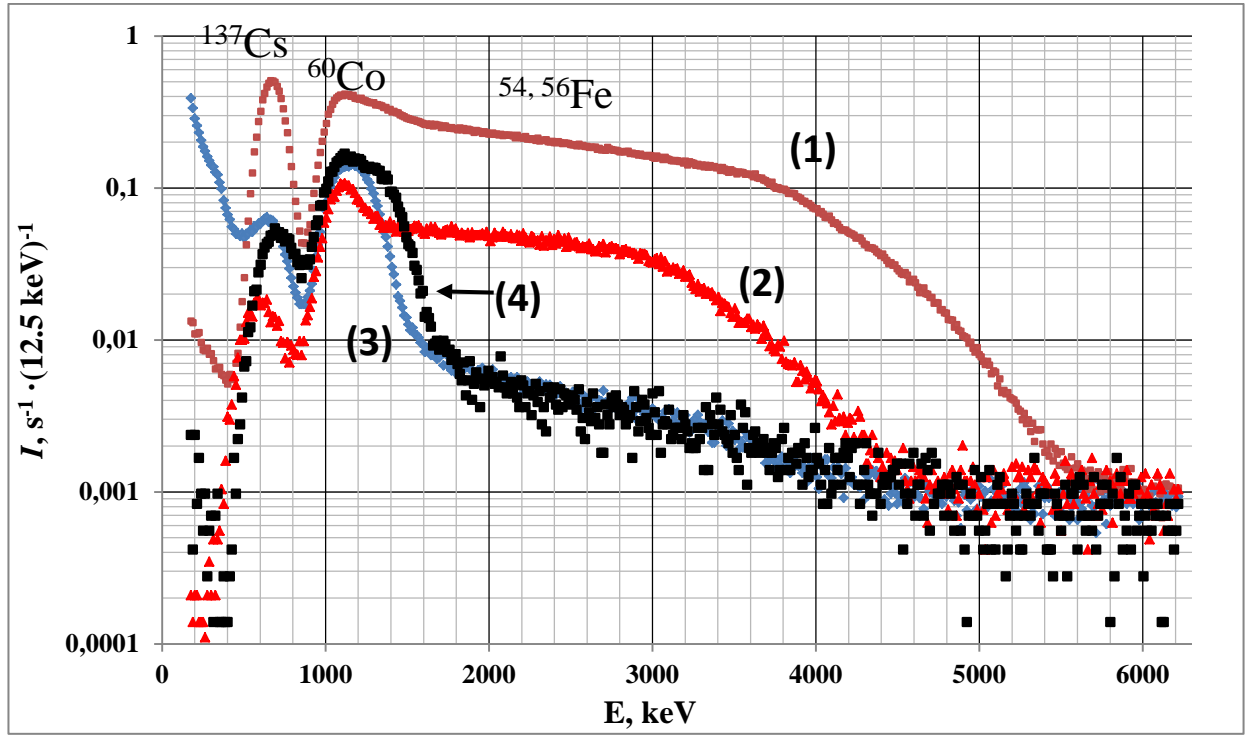

Fig. 6. Distribution of gamma-radiation intensity on the roof of the passive shielding at the distances of 1 m (curve 1) and 7 m (curve 2) from the reactor wall, as well as inside a cabin at the distance of 0.5 m (curve 3) from the reactor wall. Reactor power P=90 MW. Spectrum 4 is recorded when the reactor is off, at the distance of 9 m from the reactor wall, on the building floor.

Fig. 6 shows that there is much higher background from the instant gamma-radiation of iron on the roof of the passive shielding during the reactor operation. The further from the reactor, the less count of gamma-quanta of iron is. Measurements were made on the roof of the passive shielding of a neutron detector at the distances of 1 m, 4 m, 7 m, respectively, count rate in energy range of gamma-quanta from reaction (n, γ) on iron was 35.7 s$^{-1}$, 13.7 s$^{-1}$, 6.7 s$^{-1}$. However, inside the passive shielding at the distance of 0.5 m from the reactor wall the count rate in the same energy range was 1.1 s$^{-1}$. At the reactor switched off the spectrum was recorded on the building floor at the distance of 9 m from the reactor wall.

Fig. 7 presents the gamma spectrum shape inside the passive shielding for different distances along the way of the neutrino detector: 6.28 m, 8.38 m and 10.48 m. No noticeable alterations in the spectrum shape are observed. Moreover, for comparison gamma-spectra were measured at the reactor on and off inside the passive shielding at the point nearest to the reactor. Considerable difference in spectra was not found.

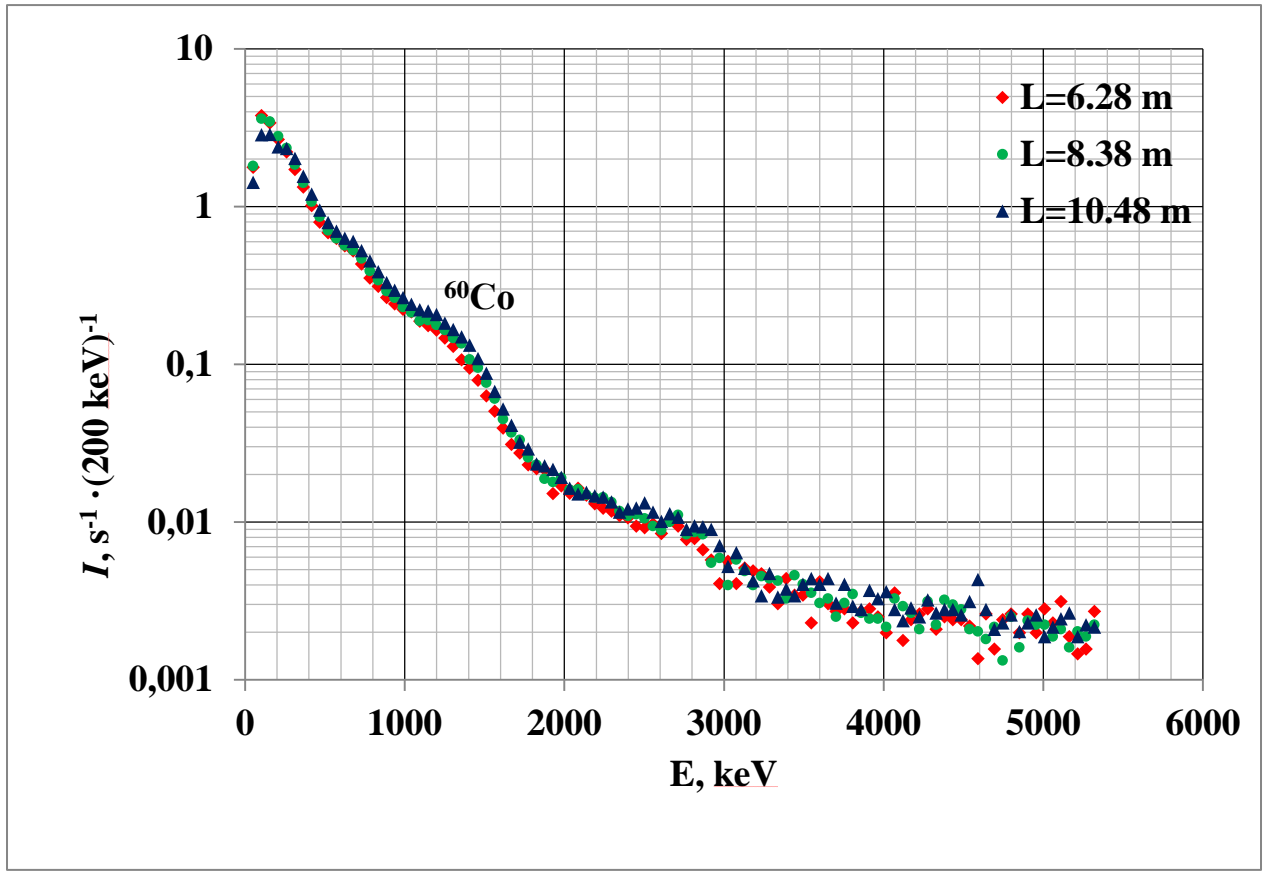

Fig. 7. Gamma-radiation spectra at detector dislocation. Reactor power is 90 MW. L is the distance from the reactor core center: - 6.28 m, - 8.38 m, - 10.48 m.

**4.2. Estimations of fluxes of fast and thermal neutrons in the neutron laboratory building and inside the passive shielding of the neutron detector**

In 2013 at SM-3 reactor a neutron laboratory building was completed for exploitation and handling equipment for the passive shielding of the neutron detector was mounted. The slide valve for the former neutron channel was carefully plugged. As a result, a flux of thermal neutrons in the neutron laboratory building decreased 29 times to the level of $(1 \div 2) \cdot 10^{-4}$ n/cm$^2$·s. This level is determined by cosmic radiation neutrons and, practically, is independent of the reactor operation. Measurements of thermal neutron fluxes were made with $^3$He detector, which represents itself a proportional counter 1 m long with the diameter of 3 cm. For registration of fast neutrons one used the same proportional $^3$He detector, but it was put into the shielding made of polyethylene (thickness of layer is 5 cm), which in its turn was wrapped in a layer of borated rubber (3 mm thick, containing 50% of boron). Thus, all background thermal neutrons before hitting on $^3$He counter are absorbed by borated rubber, while fast neutrons passing through the rubber are thermalized by polyethylene up to energy E=0.025 eV and registered by $^3$He detector. Extinction coefficient of thermal neutrons of the borated rubber was about 400 times, which enabled to separate reliably fast neutrons from thermal ones. For changing count rate (s$^{-1}$) of proportional $^3$He detector into density units of fast neutron flux (cm$^{-2}$·s$^{-1}$), a newly created fast neutron detector was calibrated according to recordings of a conventional one МКС АТ6102. For this purpose both detectors were placed side by side at the distance of 3 m from a neutron source (Pu-Be) without a thermalizer. As a result, one obtained a coupling coefficient for a count rate of the detector and of fast neutron flux $F^{fast}(n \cdot cm^{-2} \cdot s^{-1}) = 0.033 \cdot N^{fast}(s^{-1})$ for the neutron flux orthogonal to detector axis. For calibration of $^3$He thermal neutron counter one applied a similar technique with a conventional detector of thermal neutrons MKC-AT6102 and a neutron source (Pu-Be) with water moderator. Coefficient of coupling between a thermal neutron flux and a detector count rate is given by the following equation $F^{th}(n \cdot cm^{-2} \cdot s^{-1}) = 0.013 \cdot N^{th}(s^{-1})$. Made in this way, detectors of thermal and fast neutrons have sensitivity nearly two orders of value higher than that of conventional devices. They were employed for conducting low-background measurements in a neutrino laboratory. Estimations of neutron background were made before slide valve upgrading of the former neutron beam, then after upgrading and finally, after installing a passive shielding for a neutrino detector. A detector of fast neutrons was located on the roof of the passive shielding and directly on the reactor wall, i.e. at the distance of 5.1 m from the reactor core. Measuring results are presented in Fig. 8.

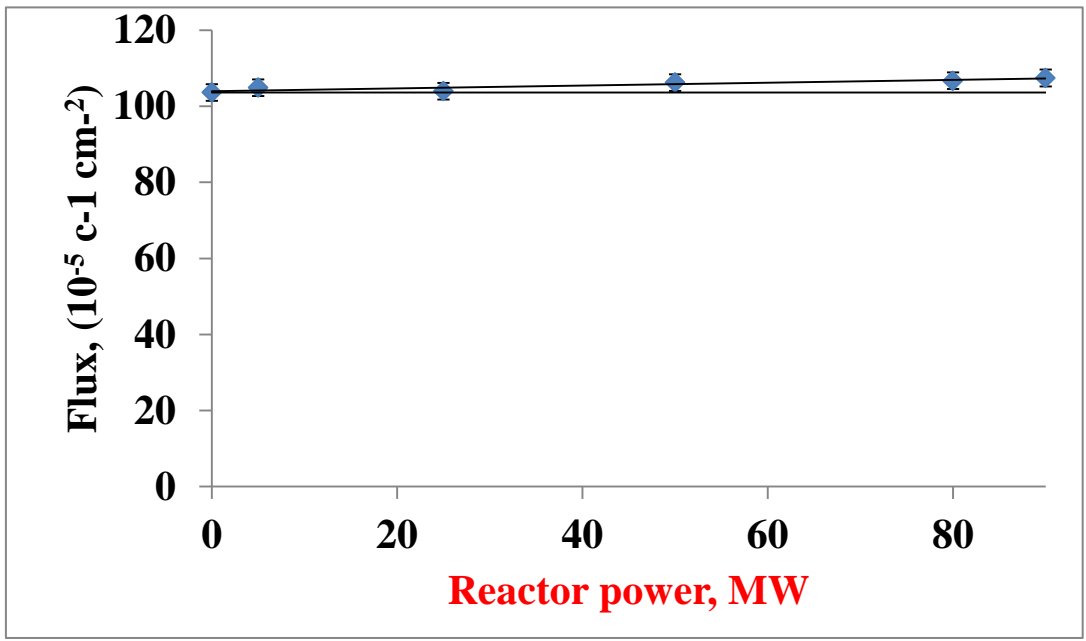

Fig. 8. Flux of a fast neutron detector at output power reactor 90MW. The counter is located on the cabin roof at the reactor wall.

From Fig. 8, one can conclude that background of fast neutrons practically does not depend on reactor power. It is determined by neutrons emerging from interaction of muons with nuclei of neighboring materials, in particular, with lead nuclei of the passive shielding. However, internal lining of the passive shielding (16 cm of borated polyethylene) suppresses this flux 12 times. Table 1 gives values of thermal and fast neutron fluxes measured inside the passive shielding and on its roof and shows factors of neutron background suppression owing to the passive shielding.

Another problem of great significance is reactor influence on the flux of fast neutrons inside the passive shielding. With this aim measurements were made of the fast neutron flux inside the passive shielding, at the nearest to the reactor wall for about ten days at the reactor on and for the same time after switching it off. When the reactor was on, the fast neutron flux was equal to $(5.54 \pm 0.13) \cdot 10^{-5}$ s$^{-1}$ cm$^{-2}$ and when the reactor was off, it was $(5.38 \pm 0.13) \cdot 10^{-5}$ s$^{-1}$ cm$^{-2}$, i.e. there was no difference within the accuracy of 2.5%.

**Table. 1**. Distribution of thermal neutron fluxes on the roof and inside the cabin of the passive shielding, with reactor power being 90 MW.

| Flux of thermal neutrons (s$^{-1}$cm$^{-2}$) | Flux of fast neutrons (s$^{-1}$cm$^{-2}$) | Measuring place |
|---|---|---|
| $(0.34\pm0.07) \cdot 10^{-5}$ | $(5.5\pm0.1) \cdot 10^{-5}$ | In the cabin |
| $(17.7\pm1.2) \cdot 10^{-5}$ | $(69\pm2) \cdot 10^{-5}$ | On the cabin roof |
| shielding factor $K_{th} = 53$ | shielding factor $K_{fast} = 12$ | |

Much more detailed measurements were made with the fast neutron detector on the cover of the neutrino one, which was moved during these measurements along the neutrino channel in the range from 6.25 m to 10.5 m. Results of these measurements at the reactor on and off are given in Fig. 9. One did not find any difference in results within statistical estimation accuracy in cases when the reactor was on or off. The background involved is determined by cosmic radiation. In these measurements the background level proved to be equal to $(8.5 \pm 0.1) \cdot 10^{-5}$ s$^{-1}$ cm$^{-2}$, which is somewhat higher than that at the reactor wall. This discrepancy can be accounted for by the detector location regarding the direction of neutron flux, i.e. vertical position at the reactor wall and horizontal one on the cover of the neutrino detector.

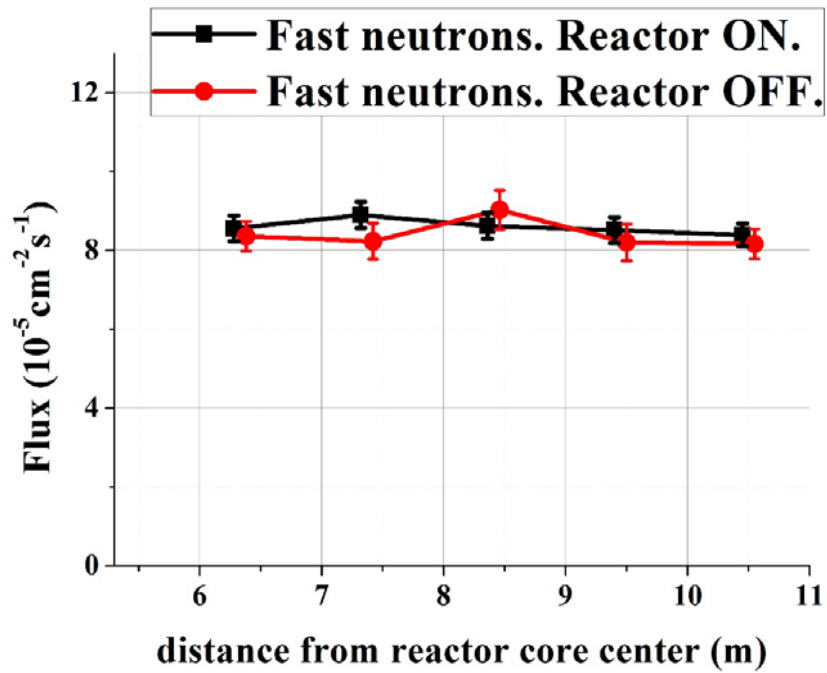

Fig. 9. Fast neutron background at different distances from the reactor core center measured with the detector of fast neutrons inside the passive shielding. The fast neutron detector was set on the cover of the neutrino detector and was dislocated together with it.

## 5. Investigation of background conditions with a neutrino detector model

### 5.1. Detector energy calibration and that of its efficiency relative to registration of neutrino events

A neutrino detector model contains 400 l of liquid scintillator BC-525 with 1 g/l Gd concentration. Description of its construction is given in Section 2. Energy detector calibration was made with $^{60}$Co gamma source and $^{252}$Cf neutron source. Activity of $^{60}$Co gamma source was 557± 55 decay/s. Every decay act emits two gamma quanta with energy 1.17 MeV and 1.33 MeV. The total absorption of the two gamma quanta in a detector results in energy release 2.5 MeV. Neutron absorption by Gd produces emission of a few gamma quanta with total energy 8 MeV. $^{252}$Cf with the total neutron flux (648± 65) n/s was used as a neutron source.

Fig. 10 shows spectra of both sources obtained as a difference in measurements with a source and without it. In this case sources were located in the detector center. They were to be moved along a vertical channel. Detector efficiency measurements were made at six height points: 1) on the detector bottom, 2) at 10 cm from the bottom, 3) at 20 cm from the bottom, 4) at 30 cm from the bottom, 5) at 40 cm from the bottom, 6) at 50 cm from the bottom or on the detector surface.

Fig. 11 gives registration efficiency values depending on the source location for Co and Cf source with the registration threshold of 0.5 MeV. In terms of these measurements one can estimate average registration efficiency of neutrino events. For the registration threshold of 0.5 MeV it is about 50%, and for the registration threshold of 3 MeV it is nearly 15%. Monte-Carlo calculations provide similar registration efficiency. We did not set ourselves the task of accurate determination of the registration efficiency of neutrino events because of our task to observe relative changes at a different distance.

To determine neutron registration efficiency with a detector, one used neutron-activity data of source $^{252}$Cf, and also a well known fact that one fission act $^{252}$Cf emits 3.767(4) neutrons. Neutron registration was made by means of isolation of correlated events, with the start being a signal at the moment of nucleus fission, i.e. from gammas and fast neutrons. Stop signal was gamma-quanta registration at neutron capture by Gd. Electronic devices of the installation enabled recording multiple stops at one start. Thus, the ratio of the mean number of registered neutrons to that of emitted neutrons (3.767) corresponds to the registration efficiency of one neutron. A similar technique for determining neutron efficiency registration was employed in an experiment at WWR-M reactor, however, as the start signal one used a signal from fission fragments recorded with a semiconductor detector. Fig.12a shows energy spectra of start and stop signals in an experiment with $^{252}$Cf source. The integral under the curve of stop signals is 2.75 times bigger than that under the curve of start signals, which corresponds to 73% registration efficiency of one neutron.

Fig.12b shows the time spectrum of neutron registration. One can see that neutron thermalization process to thermal velocity is 5 -7 µs, while thermal neutron life time in a scintillator with Gd is 31.3 µs.

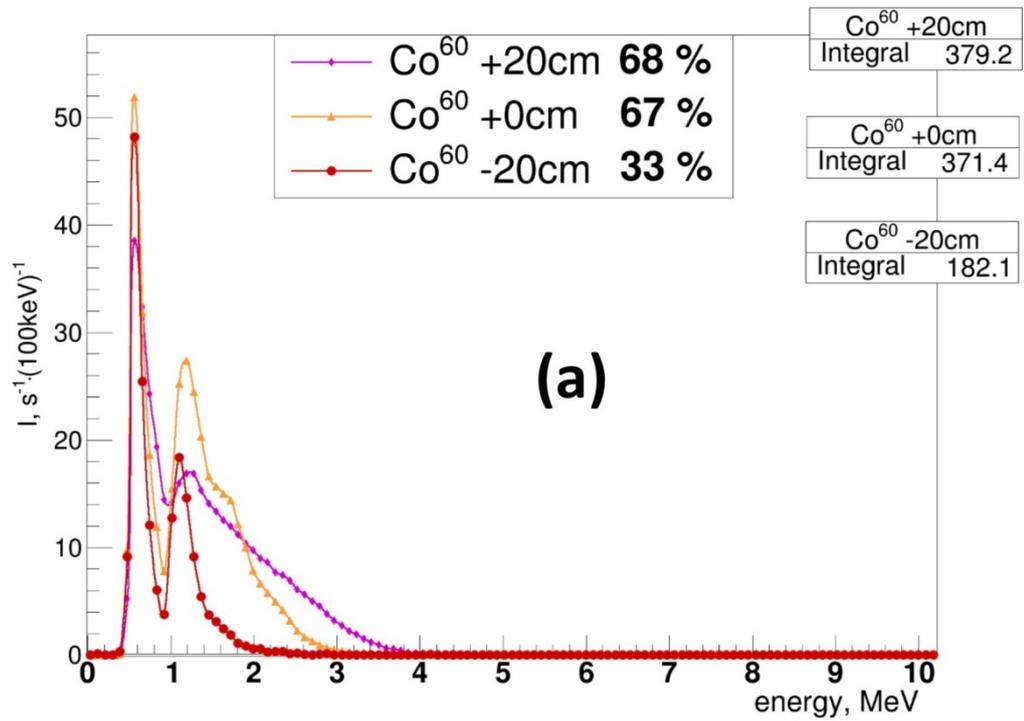

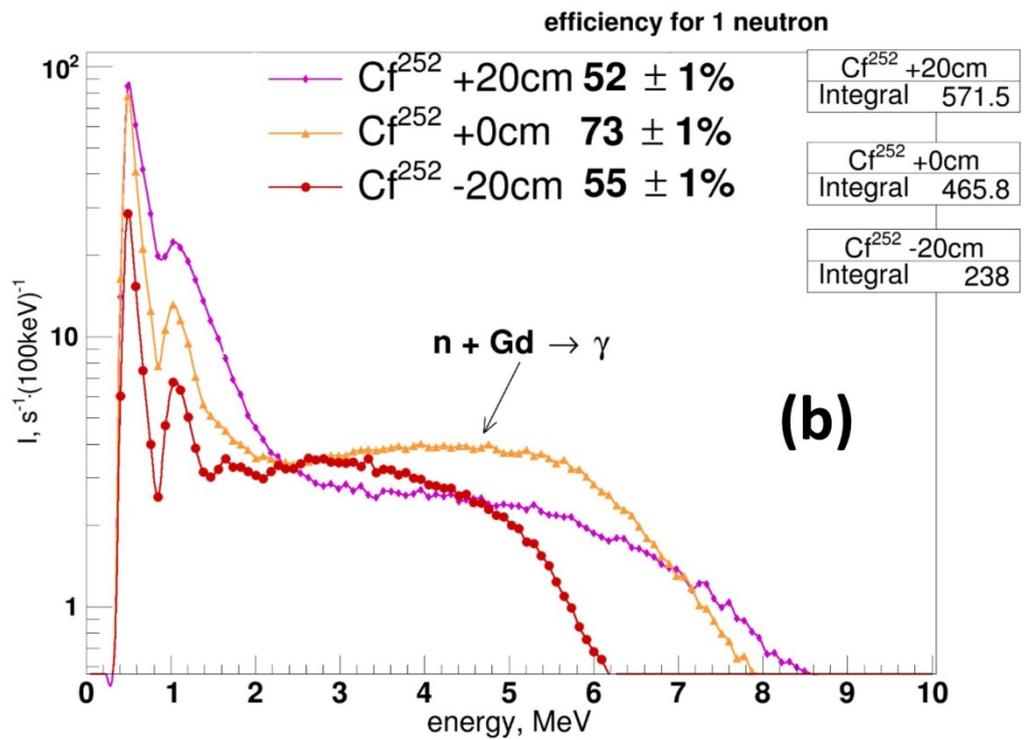

Fig. 10. Calibration of a neutrino detector model with gamma radiation source ($^{60}$Co) and neutron source ($^{252}$Cf): a) energy spectrum of $^{60}$Co, b) energy spectrum of $^{252}$Cf.

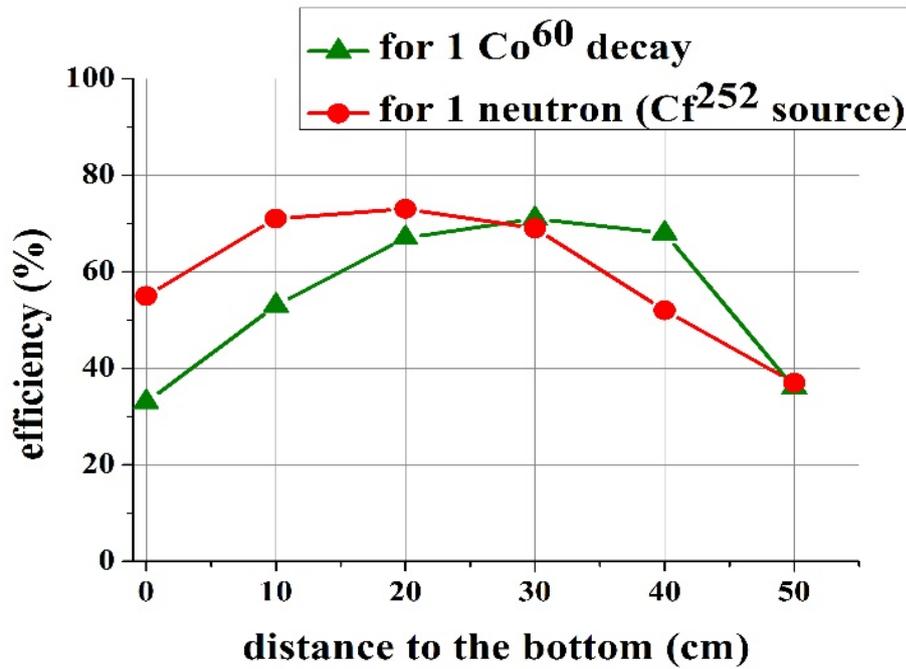

Fig. 11. Registration efficiency of neutrons and gamma-quanta from sources ($^{252}$Cf and $^{60}$Co correspondingly) depending on their location.

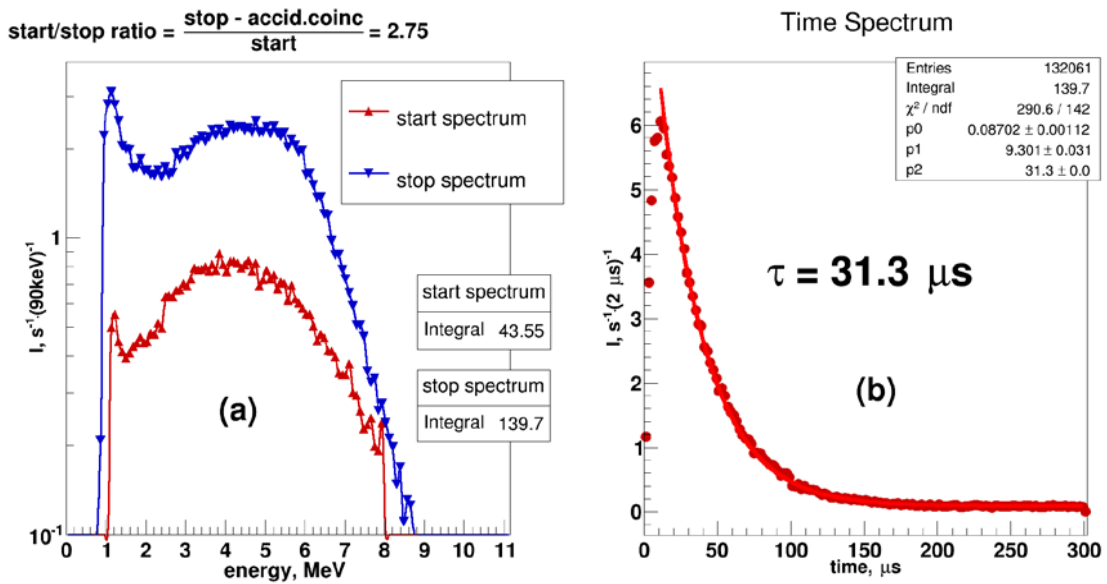

Fig. 12. a) Energy spectra of start and stop signals in Cf source experiment, b) time spectrum of neutron registration, neutron thermalization process to thermal rate is 5 -7 µs, while thermal neutron life time in the scintillator with Gd is 31.3 µs.

### 5.2. Investigation of cosmic ray background

The neutrino detector model involved can be used for cosmic ray registration and gamma quanta from radioactive contaminations. Fig. 13 presents the neutrino detector model spectrum, which can be conventionally divided into 4 parts.

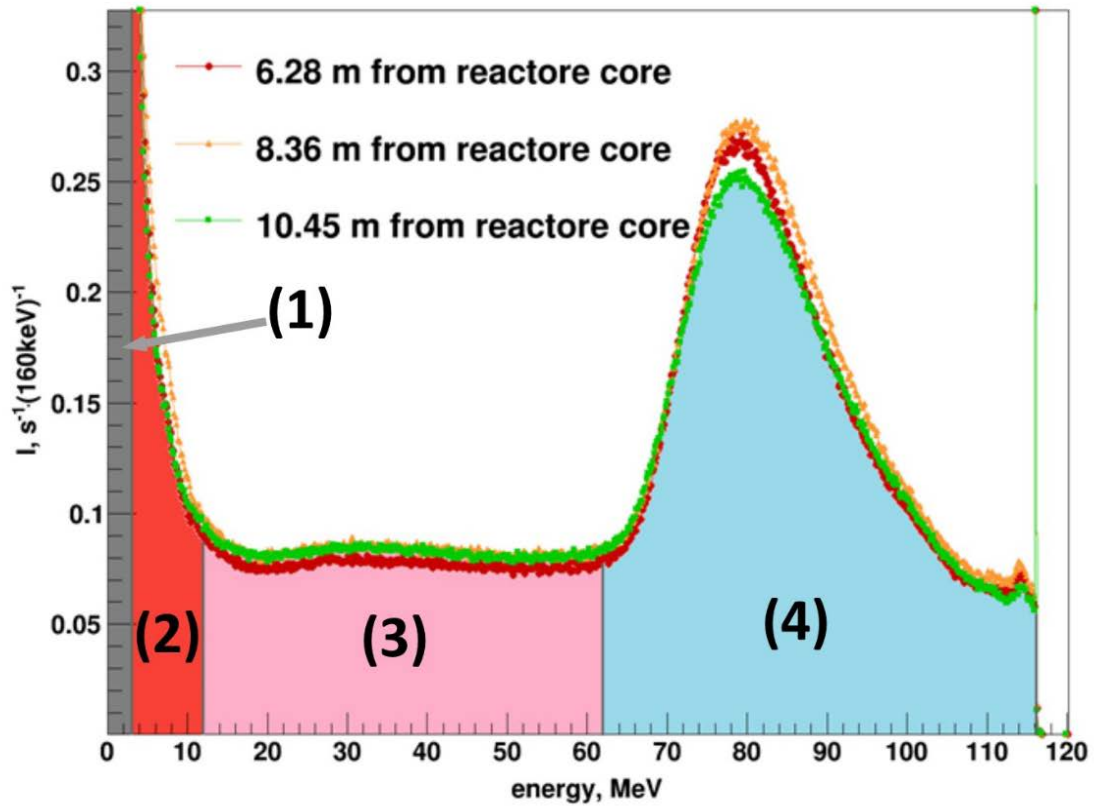

Fig. 13. Detector energy spectrum at different distances from reactor core and conventional division of spectrum into zones: 1 – radioactive contamination background, 2 - neutrons, 3 – cosmic radiation soft component, 4 -muons.

The first part of it up to 2 MeV is relevant to radioactive contamination background, second one from 2 MeV to 10 MeV covers the registration neutron area, since it corresponds to gamma-quanta energy at neutron capture by Gd. The range from 10 to 60 MeV is relevant to cosmic radiation soft component produced by muon decay and muon capture in substance. And finally, the range from 60 – 120 MeV is related to muon component passing through the detector. Here are also shown small alterations of spectrum shape for different detector positions.

Fig. 14 shows the intensity dependence on distance along an experimental base for different spectrum parts on a big scale. Dependence on the distance for the spectrum part associated with radioactive contamination can be accounted for by previous exploitation of the building. The highest activity is concentrated near the gate device and on the floor along the rails. It should be noted, that in estimations with detector NaJ a similar dependence was not discovered, as NaJ detector was situated at the distance of 1.4 m from the floor and was screened by the detector mass itself. A neutrino detector keeps recording radioactive contamination on the floor, when moving along the rails. For other spectrum parts of cosmic background the distance dependence with variation of about 10% are also observed. Such dependences seem to be associated with distribution of concrete structure in the building (Fig. 14).

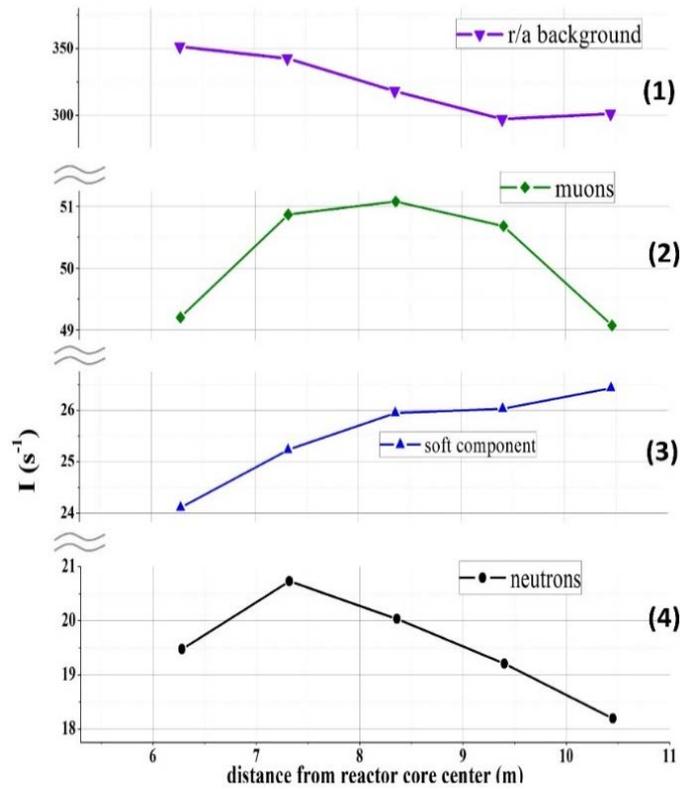

Fig. 14. Dependence of the detector count rate in different ranges of energy spectrum of a neutrino detector model (enlarged scale with discontinuities) 1 – radiation contamination background, 2 - neutrons, 3 – cosmic radiation soft component, 4 - muons.

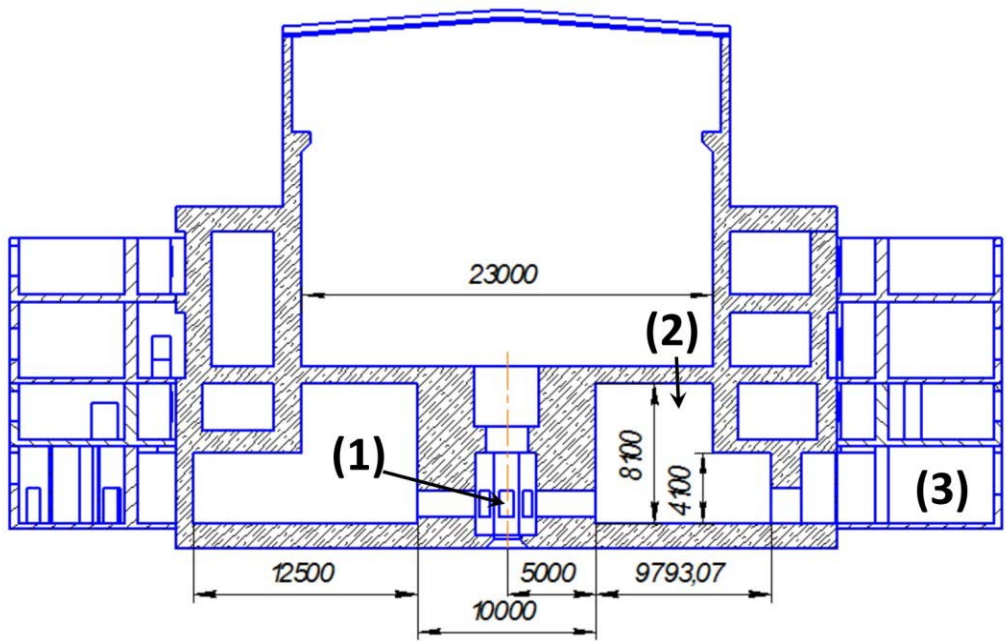

Fig. 15. Section of the reactor building. 1 – reactor, 2 – neutron laboratory, 3 – computer room.

In the course of long-term measurements temporary variations of cosmic radiation intensity were found. They are caused by fluctuations of atmospheric pressure and temperature drift during season changes. It is a well known barometric and temperature effect of cosmic rays [7-9]. Muons are formed in the upper layers of atmosphere. Higher pressure gives rise to a larger amount of substance over the detector and to intensity attenuation of cosmic rays. Fig. 16 shows anti-correlation effect between atmospheric pressure and total intensity of rigid and soft components of cosmic radiation, i.e. within energy range from 10 to 120 MeV. This effect is barometric. Behavior of rigid and soft components is distinguished by presence of additional long-term drift, with the drift sign being opposite for rigid and soft components. It is the so called temperature effect which is interpreted in the following way. At temperature rise of the lower atmospheric layers, their expansion results in raising the height of the layer forming muon fluxes. Because of increase in the way to Earth the share of the decayed muons increases. Thus, rigid component intensity (muons) decreases and soft component intensity (decay products: electrons, positrons, gamma quanta) rises. Fig. 17 shows a drift effect with opposite signs for rigid and soft components of cosmic radiation at a higher temperature of the lower layers of atmosphere in the vicinity of the Earth surface from January till April from –30°C to +10°C.

As a result of studying background conditions for performing an experiment on search for neutrino oscillations at short distances, it became clear that background conditions were extremely unfavorable. Cosmic background depends on the distance from the reactor core center because of distribution of concrete structure of the building. Moreover, cosmic background is changing with time due to atmospheric pressure and temperature fluctuations in lower atmosphere layers. However, the following ways of overcoming these problems can be suggested. Firstly, one can implement monitoring of cosmic ray intensity regarding a higher energy part of the detector spectrum starting with 10 MeV. Applying monitoring data, one can introduce corrections for a cosmic ray background. Secondly, measurements of the distance dependence should be made by means of scanning the distance in the moving mode from one point to another every 45 - 60 min. Thus, the total range is measured for 4 - 5 hours, which is less than the time of atmospheric pressure variation. It enables to average considerably the time variation effect of cosmic rays.

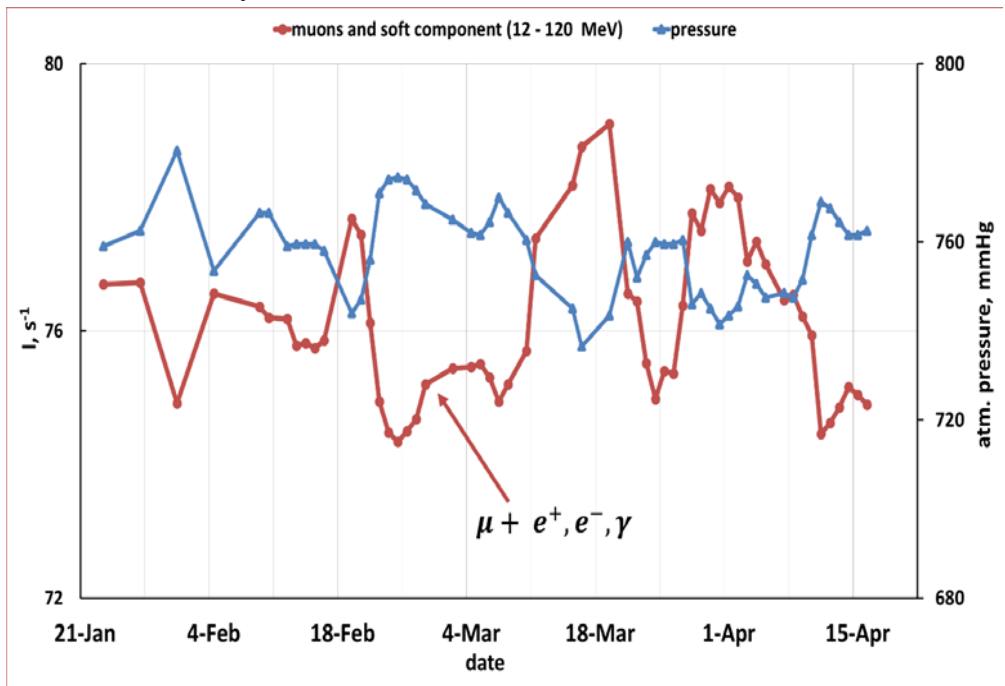

Fig. 16. Barometric effect of cosmic rays: the left axis shows summary detector count rate in the area 3 and 4, the right axis shows atmospheric pressure, horizontal axis gives the measurement time since 23d of January to 15th April 2014.

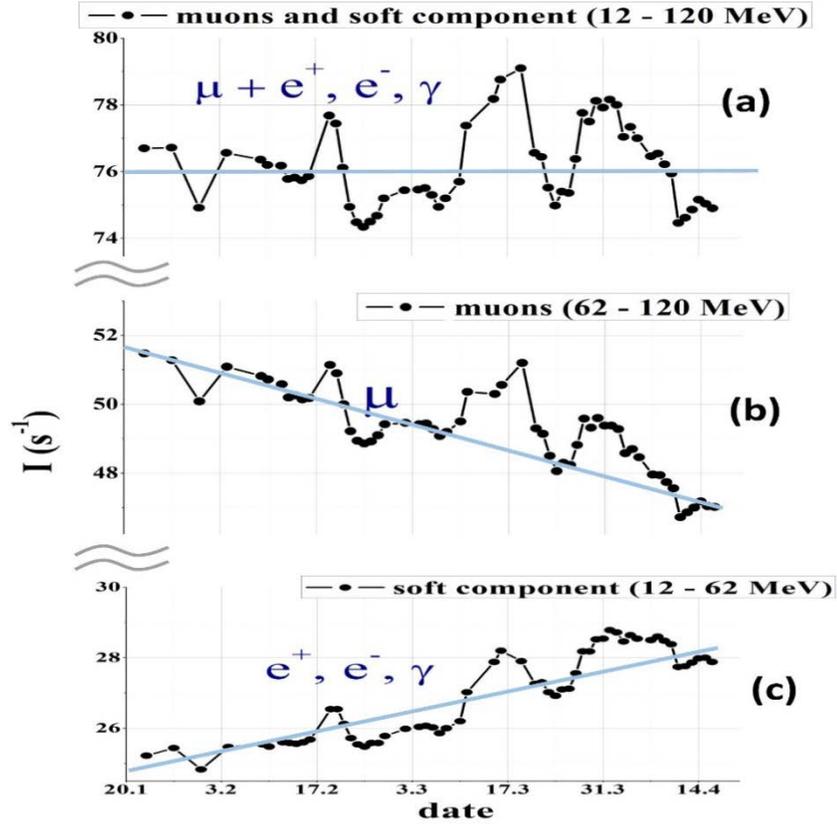

Fig. 17. Barometric and temperature effects of cosmic rays; a) summary detector count rate in the area 3 and 4, b) detector count rate in the area 4, c) detector count rate in the area 3. Along the horizontal axis is the measurement time since 23d of January till 15$^{th}$ of April 2014.

## 6. Testing measurement of antineutrino flux from SM-3 reactor and its distance dependence.

### 6.1. Energy and time spectra of correlated events

As it was noted earlier, in measuring antineutrino flux from the reactor the technique of correlated coincidences is employed for distinguishing the registration process of antineutrino - $\tilde{\nu}_e + p \rightarrow e^+ + n$. Fig. 18 gives the spectrum of delayed coincidences. The background of random coincidences is subtracted. One can see two exponents (straight lines in logarithmic scale), which correspond to a muon decay and a neutron capture by Gd. Without employing an active shielding, the integral under the first exponent corresponds to the muon stop rate 1.54 µ/s, and the exponent (2.2 µs) corresponds to a muon lifetime. The integral under the second exponent is relevant to the neutron capture rate in the detector – 0.15 n/s, and the exponent (31.3 µs) corresponds to neutron lifetime in the scintillator at 0.1% Gd concentration.

The number of muon stops per second corresponds to estimation on the muon flux and scintillator mass calculation, while the number of captured neutrons per second corresponds to the calculated rate of neutron formation in the detector itself, caused by a muon flux passing through it. It points out that, in general, one succeeded in solving the task under consideration by means of the passive shielding in combination with lead placed outside with 16 cm of borated polyethylene inside. Indeed, addition of 10 cm borated polyethylene upon the detector cover did not alter the neutron capture rate in the detector. Use of a ban from active shielding and the detector, which gives evidence for muon passing, allows to suppress the capture rate by the detector to level $1.8 \cdot 10^{-2}$ n/s. Fig. 18 presents the first version of the active shielding. Detailed studies of the active shielding are quoted in the next section.

Fig. 19 shows energy spectrum of stop signals within the time range from 0 to 10 µs and that of stop signals within 10 to 100 µs. One can make the following conclusions from the obtained data. Choosing the time interval from 0 to 10 µs, we highlight a process of muon stops in the detector and it is seen, that spectrum of decay products covers the range from 0 to 50 MeV, i.e. it corresponds to continuous spectrum of electrons (positrons) and gamma quanta at muon decay. In selecting the time interval of delayed coincidences from 10 to 100 µs, one obtains spectrum of stop signals in the range from 2 to 8 MeV, which corresponds to gamma radiation spectrum at neutron capture by Gd. Horizontal part of spectrum over 10 MeV is related to background of accidental coincidences.

The most essential experimental problem is possibility of distinguishing correlated events against background of accidental coincidences. Fig. 20 presents examples of correlation signal measurements.

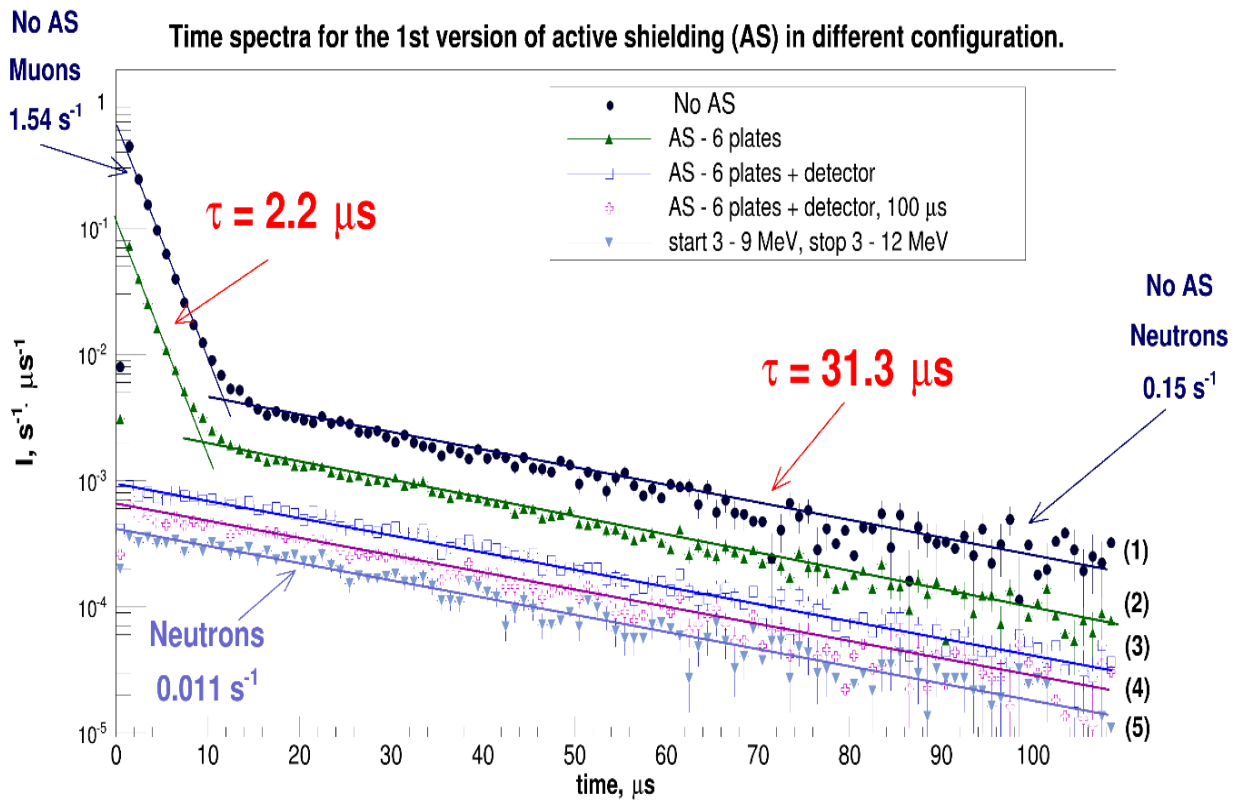

Fig. 18. Time spectra at different configurations of the active shielding: 1 - no active shielding, 2 – plates of the active shielding are on, 3 – the same + ban from the detector at signals higher than 12 MeV, 4 – the same + ban on 100 µs after the detector signal, at energy higher than 12 MeV, or after the signal in the active shielding, 5 – the same + limit on start and stop signals in ranges 3 – 9 MeV and 3 – 12 MeV respectively.

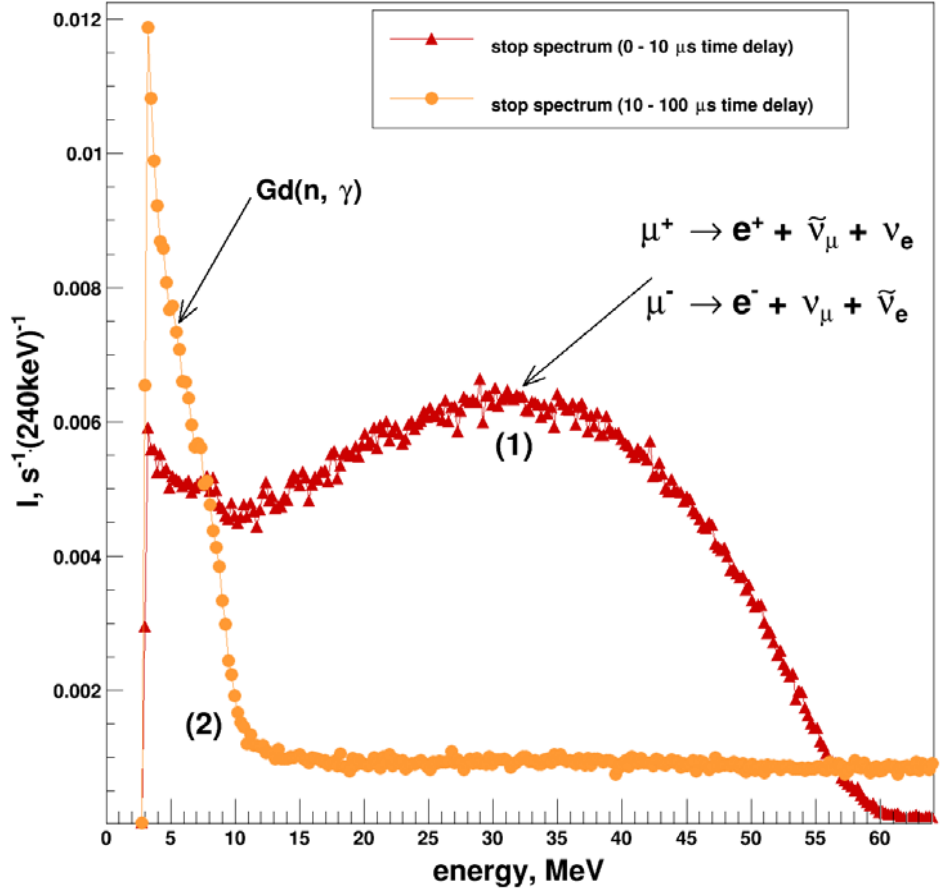

Fig. 19. Energy spectra of time delay signals in the range of 0 – 10 µs (1) and 10 – 100 µs (2).

Measurements were made for 300 µs, the last 100 µs used for measuring background of accidental coincidences. At lower threshold of start and stop signals 3 MeV, contribution of accidental coincidences is very small (Fig. 20a). In decreasing the lower threshold the number of accidental coincidences increases, however, the number of correlated events grows (Fig. 20b). Nevertheless, accuracy of correlated signal is not growing. Radioactive contamination background remains high enough. Unfortunately, its reduction does not seem to be possible for the time being. However, it should be noted, that a much more essential problem appears to be concerned with correlated background related to cosmic radiation, i.e. muons and fast neutrons.

Expected count rate of neutrino events is calculated with the following ratio:

$$n_\nu = \left(\frac{W_{eff}}{E_{eff}}\right) \cdot (4\pi R^2)^{-1} N_p \cdot \sigma_f \cdot \varepsilon , \qquad (2)$$

where $W_{eff}$ – reactor thermal power, $E_f$ – energy release per fission, including direct and delayed processes, $R = 7.07\ м$ – distance from the reactor core to the detector center, $N_p$ – the number of protons in the detector, $\sigma_f$ – efficient neutrino cross-section per fission. The data from work [10-11] were used for calculations. The number of expected neutrino events at reactor power 90 MeV at the distance of 7.07 m is $0.67 \cdot 10^{-2}$ n/s × ε, where ε – detector efficiency on registration of occurring neutrino events according to reaction $\tilde{\nu}_e + p \rightarrow e^+ + n$.

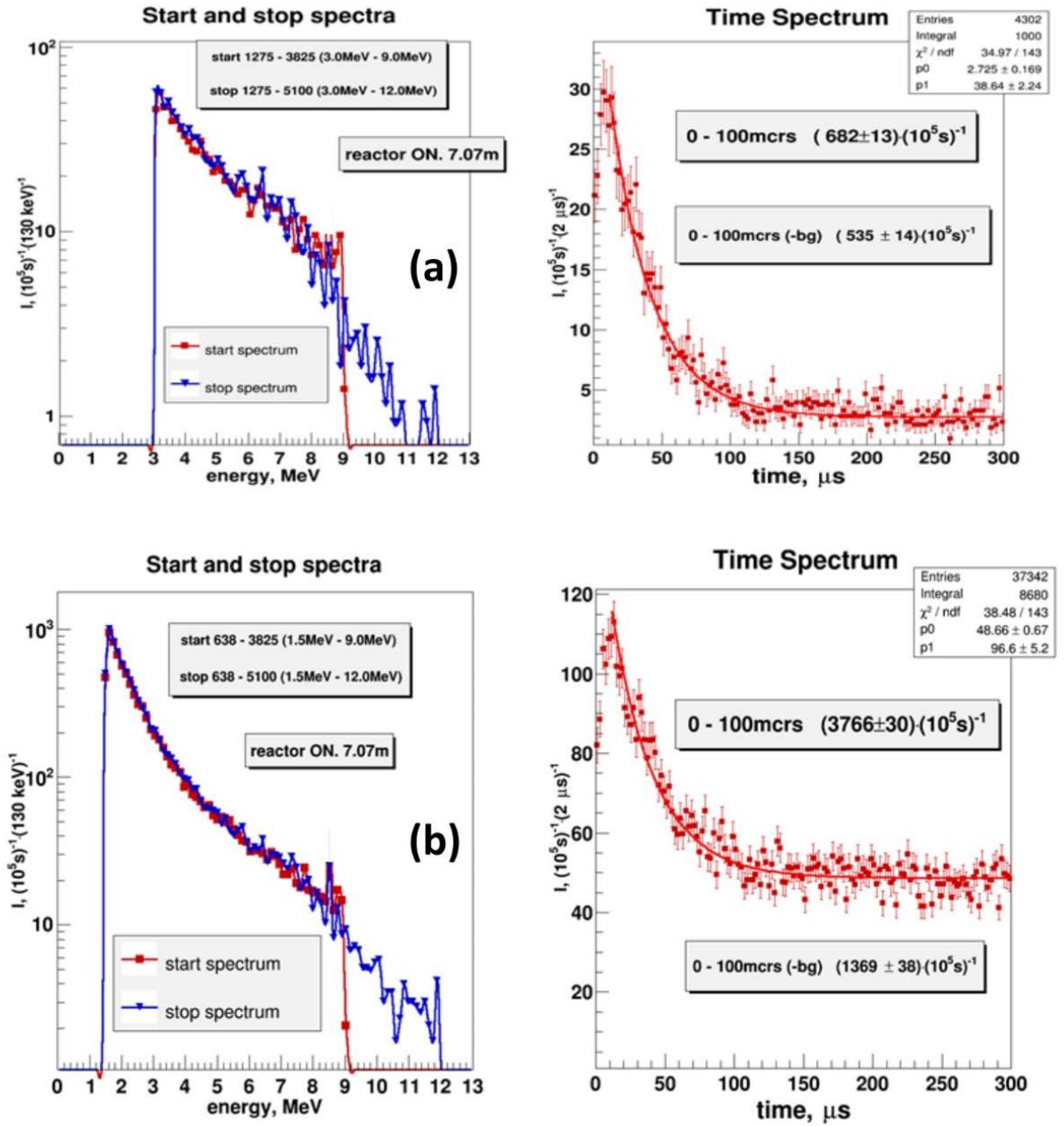

Fig. 20. Energy spectra of direct and delayed signals and time spectra: a) threshold of start and stop signals 3 – 9 MeV and 3 – 12 MeV respectively, b) 1.5 – 9 MeV and 1.5 –12 MeV.

Experimental number of registered neutrino events is determined as a difference at the reactor on and off. Fig. 21a gives the ratio of registered neutrino events to the total number of neutrino events depending on thresholds of start and stop signals for the detector position 7.07 m. For comparison Fig. 21b presents Monte-Carlo calculation. At the lower 1.5 MeV threshold of start and stop signals registration efficiency of the reactor antineutrino is about 50%, which does not disagree with Monte-Carlo calculation and an experimental estimation of neutron registration efficiency in an experiment with Cf source. Undoubtedly, estimation accuracy is rather approximate, i.e. on ±15% level.

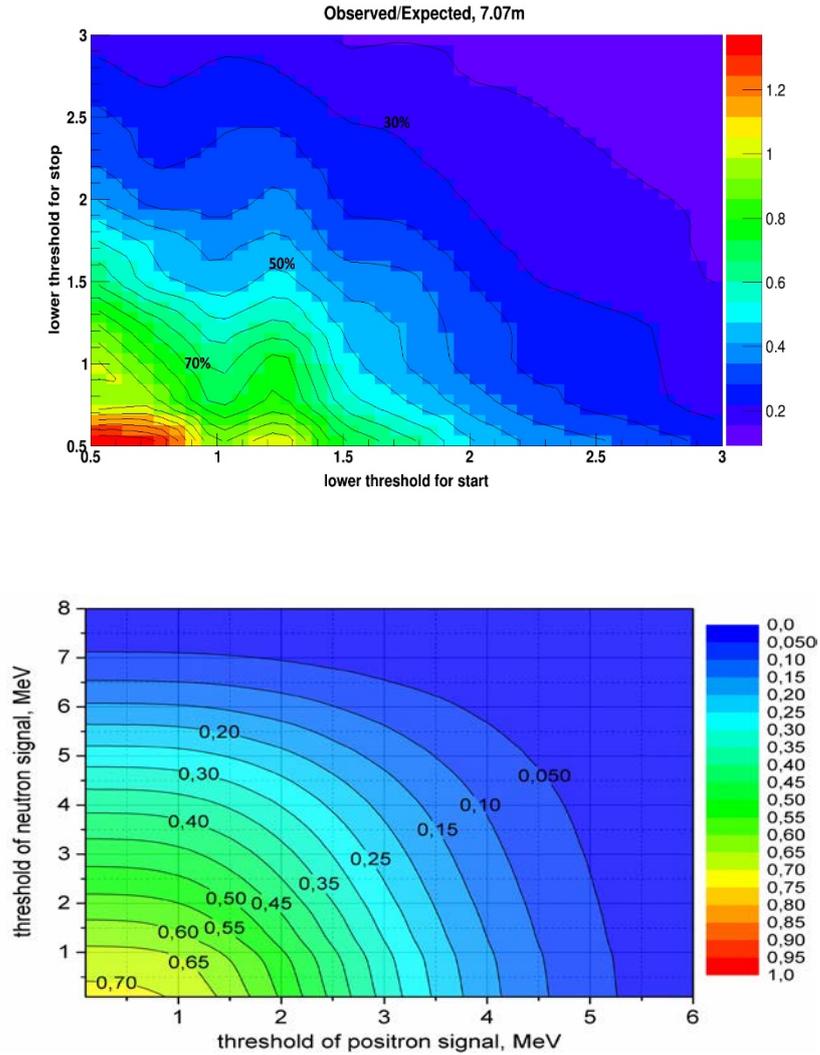

Fig. 21. a) Experimental efficiency on registration of antineutrino events depending on lower thresholds of registration of start and stop signals, b) registration efficiency of antineutrino events from Monte-Carlo calculations depending on lower registration thresholds of start and stop signals.

**6.2. Active shielding of the detector**

The next stage of research was concerned with the active shielding of the detector. The first part of measurements was made with the first version of the active shielding, with scintillator plates being 3 cm thick. Its application resulted in reducing the number of muon stops by 7 times to the level 0.27 μ/s, with the number of neutron captures being reduced by 2 times to the level 0.078 n/s (Fig. 18). The next stage of suppressing correlated neutron events was to limit the energy of start signals from the detector to the level of 12 MeV. After that the count rate of neutron events dropped to $2.7 \cdot 10^{-2}$ n/s. It turned out that it is reasonable «to lock» the detector for another 100 μs after the signal emergence in the detector at energy over 12 MeV or after the signal in the active shielding. This decreased count rate of correlated neutron captures in the detector to the level of $1.8 \cdot 10^{-2}$ n/s. Finally, restriction of start and stop signals to the ranges 3–9 MeV and 3–12 MeV, respectively, decreased the neutron capture rate to the level of $1.1 \cdot 10^{-2}$ n/s, with the number of expected neutrino events noticeably dropping.

The second version of the active shielding was made of plates 12 cm thick. In this case one succeeded in obtaining the effect-background ratio equal to 0.23 rather than 0.12 for the point nearest to the reactor (Fig. 22). As earlier this ratio 0.23 remains unsatisfactory to measure within a few percent accuracy dependence $1/R^2$, i.e. for search for neutrino oscillation. It is necessary to obtain the ratio equal to unit for the furthest point from the reactor. We assume the remaining background of correlated events to be mainly related to fast neutrons of cosmic background. Fast neutron produces a start signal via a recoil proton and gives a stop signal at absorbing the same neutron. Now we are undertaking investigation on separating signals according to an impulse shape, as a signal shape in recording a recoil proton or a positron must be different. We do hope to improve the signal-background ratio by increasing the detector efficiency due to enlarging the detector volume. The full scale detector volume is expected to be increased by 4 times.

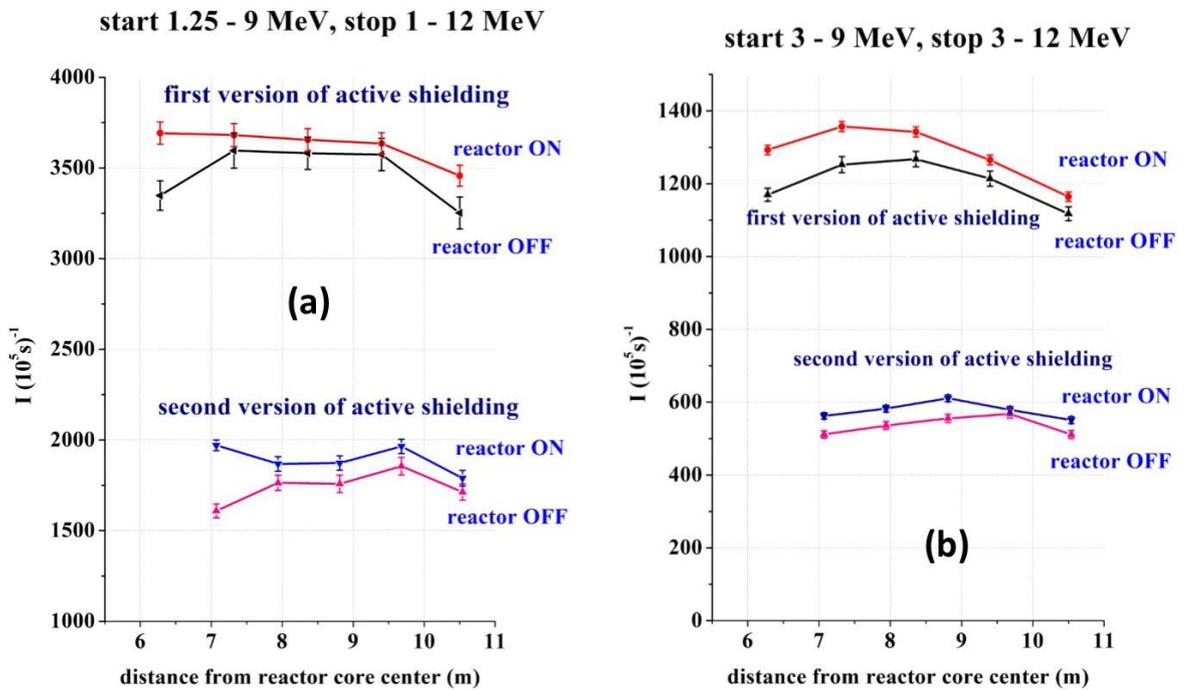

Fig. 22. a) measuring results on count rate of correlated events for start signals within the energy range of 1.25 – 9 MeV and stop signals 1 – 12 MeV at the reactor on and off, as well as for two versions of the active shielding, b) results of the same measurements for start signals within the energy range of 3 – 9 MeV and for stop signals 3 – 12 MeV.

To conclude, investigations with the external active shielding were carried on (external one with respect to the passive shielding). On the roof of the passive shielding over the detector was installed the active shielding («umbrella») made of scintillator plates 12 cm thick and the total area 2 x 3 m. Taking into account the fact that the detector area is 0.9 x 0.9 m, such an «umbrella» must capture the main muon flux flying into the neutrino detector area. As a result, it improved the effect-background ratio by 15% only and the ratio increased up to the level of 0.32. The remaining correlated background is likely to be related to fast neutrons which are only partly blocked by the active shielding. Thus, we mainly hope on employing the technique of signal separation according to the impulse shape.

## 6.3 Measuring the dependence of reactor antineutrino flux on the distance from the reactor core center

At the next stage measurements were made of antineutrino flux from SM-3 reactor and its distance dependence. Measuring results of correlated signals depending on distance with the reactor on and off were presented in Fig. 22. From the difference of these results was derived the dependence of the reactor antineutrino flux on the distance from the reactor core center (Fig. 23).

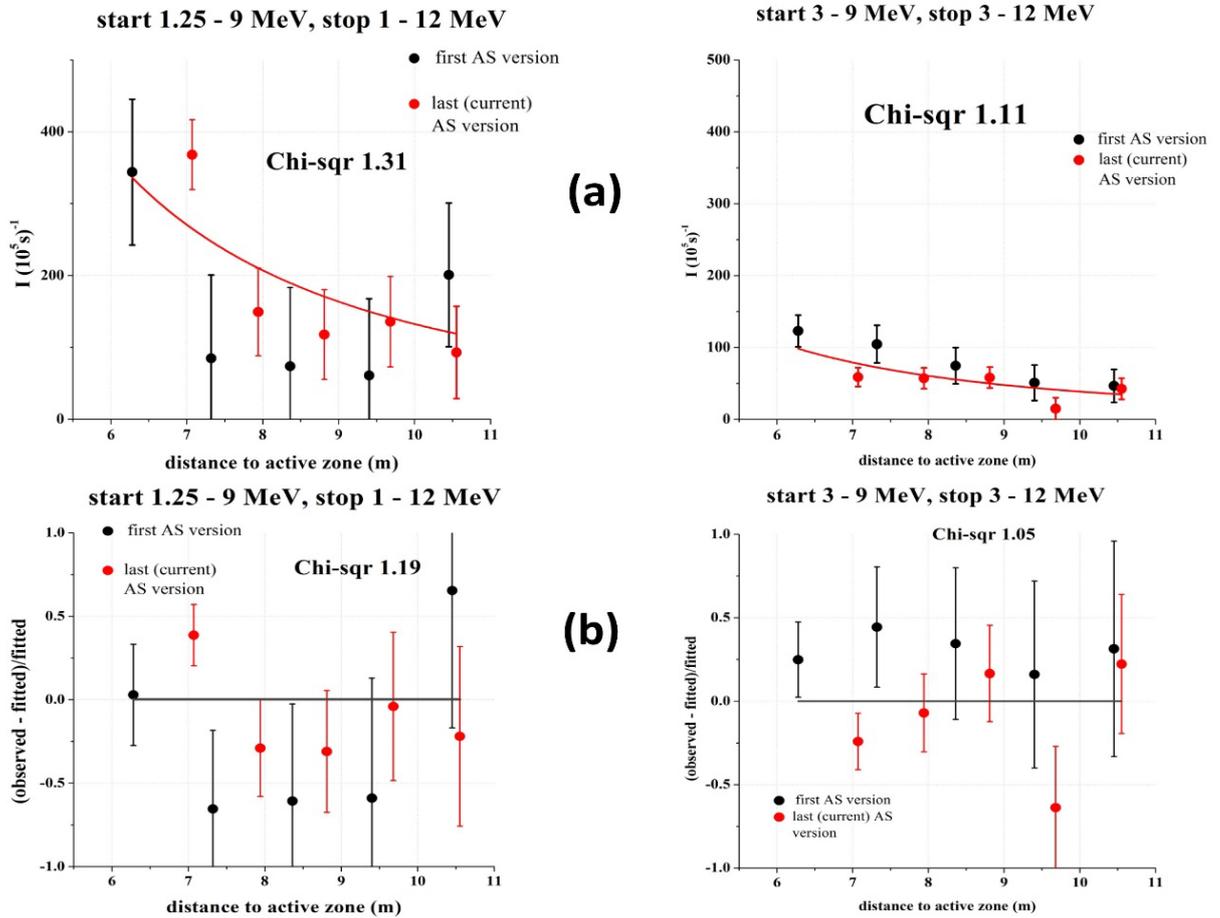

Fig. 23. a) On the left - dependence of count rate difference of correlated events (reactor on – reactor off) on distance from the reactor core center within the range of 1.25 – 9 MeV and for stop signals 1 – 12 MeV, on the right – the same for the energy range of 3 – 9 MeV and 3 – 12 MeV; b) – treatment of the same data on deviation from the law $1/R^2$.

Unfortunately, attempt to increase statistics for the sake of a wider energy interval does not improve the situation, since in a small energy area the contribution of correlated events from cosmic background is growing.

In order to verify that the difference effect is mainly relevant to antineutrino of reactor SM-3, additional measurements were made, when another lining of borated polyethylene of 0.3 m was installed near the reactor wall. It could attenuate the flux of fast neutrons from the reactor 3-4 times. Measuring results with an additional wall show that difference decrease (reactor on – reactor off) was not found within the statistical measuring accuracy of 20%.

## 7. Conclusions

Summarizing, the following conclusions can be done.

1. For the first time, an attempt was realized to measure the reactor antineutrino flux dependence at short distances (6 – 11 m) from the reactor core center. Undoubtedly, the accuracy is not sufficient for making conclusions concerning the statement of the task on search for a sterile neutrino. The task was only aimed at studying the possibility of performing such an experiment at the cosmic background level on the Earth surface and at the reactor operation background level. This experiment made use of a prototype of a small volume detector.

2. The main problem of this experiment is concerned with correlated background related to cosmic radiation. Cosmic background depends on the distance from the reactor core center due to the distribution structure of concrete mass of the building. Moreover, cosmic background is altering with time because of atmospheric pressure and temperature fluctuations in the lower layers of atmosphere. However, to overcome these problems the following measures can be suggested. Firstly, one can perform monitoring of cosmic ray intensity regarding the high energy part of the detector spectrum starting with 10 MeV. Secondly, measurements on distance dependence should be made by the method of scanning distance. It provides considerable averaging of the temporary variation effect of cosmic rays.

3. Employment of the active shielding allows suppressing correlated background of cosmic radiation only by 66%. This cosmic background component seems to be related to muons. It can be controlled by the active shielding. The active shielding is practically not capable of controlling the neutron component, thus, it is required that the technique of separating signals from recoil protons and positrons according to impulse shape should be applied.

The carried out work gave enough information for development of the full-scale detector. At the moment the project of the full-scale detector with a full volume of 3 $m^3$ is developed. We assume that implementation of the project and the pulse shape discrimination method will bring the effect-background ratio closer to value of 1 and will considerably increase the statistical accuracy of the experiment. It will allow starting the studies connected with the search for oscillation of reactor antineutrino at short distances.

This work was supported by the Russian Foundation for Basic Research, grant no. 14-22-03055-ofi_m.